\documentclass[prl,twocolumn,showpacs,groupedaddress]{revtex4}%
\usepackage[dvips]{epsfig}
\usepackage{subfigure}
\usepackage{float}
\usepackage{amsmath}
\usepackage{amssymb}
\usepackage{amsfonts}
\usepackage{rotating} 
\usepackage{euscript}
\usepackage{subfigure}
\usepackage{enumerate}
\usepackage{hhline}
\usepackage{supertabular}
\usepackage{multirow}
\usepackage{tabularx}

\newcolumntype{Y}{>{\centering\arraybackslash}X}

\begin{document}

\title{Rayleigh scattering, mode coupling, and optical loss in silicon
microdisks} \author{Matthew Borselli} \author{Kartik Srinivasan}
\author{Paul E.  Barclay} \author{Oskar Painter}
\email{borselli@caltech.edu} \affiliation{Department of Applied
Physics, California Institute of Technology, Pasadena, CA 91125, USA.}
\date{\today}

\begin{abstract}
High refractive index contrast optical microdisk resonators fabricated from silicon-on-insulator wafers are studied using an external silica fiber taper waveguide as a wafer-scale optical probe.  Measurements performed in the 1500 nm wavelength band show that these silicon microdisks can support whispering-gallery modes with quality factors as high as $5.2 \times 10^5$, limited by Rayleigh scattering from fabrication induced surface roughness.  Microdisks with radii as small as $2.5$ $\mu$m are studied, with measured quality factors as high as $4.7 \times 10^5$ for an optical mode volume of 5.3$(\lambda/n)^3$. 
\end{abstract}

\maketitle

Recent studies of optical resonators in glass based microspheres\cite{Vernooy,Weiss,Cai}, microrings\cite{Little1}, and microtoroids\cite{Armani} have highlighted the applications afforded by the extremely long photon lifetime of whispering-gallery-modes (WGMs) supported by these structures.  Furthermore, recent work by Ilchenko \textit{et al.} illustrated the advantages of creating WGMs in crystalline materials\cite{Ilchenko1}.  Applications for such devices include quantum networking\cite{Kimble2}, low threshold non-linear optical sources\cite{Spillane1}, and compact micro-optical circuits\cite{Little1}.  The ability to create similar high quality factor ($Q$) WGM resonators in III-V or silicon (Si) semiconductors has thus far been hampered by the large refractive index of most semiconductors and the resulting sensitivity to surface roughness\cite{Gayral,Choi}.  In this Letter we describe measurements of micron-sized Si microdisk resonators supporting transverse-magnetic (TM) WGMs with significantly reduced sensitivity to disk-edge roughness.  These modes have measured $Q$ values as high as $5.2 \times 10^5$ and effective modal volumes ($V_{\text{eff}}$) as small as $5.3$ cubic wavelengths in the material.  The largest $Q/V_{\text{eff}}$ ratio is measured to be $8.8 \times 10^4$, greater than the values measured in ultra-small volume photonic crystals\cite{Srinivasan7} and comparable to the values measured in ultra-high-$Q$ microspheres and microtoroids\cite{Vernooy,Armani,Ilchenko1}.

\begin{figure}[t]
\begin{center}
\epsfig{figure=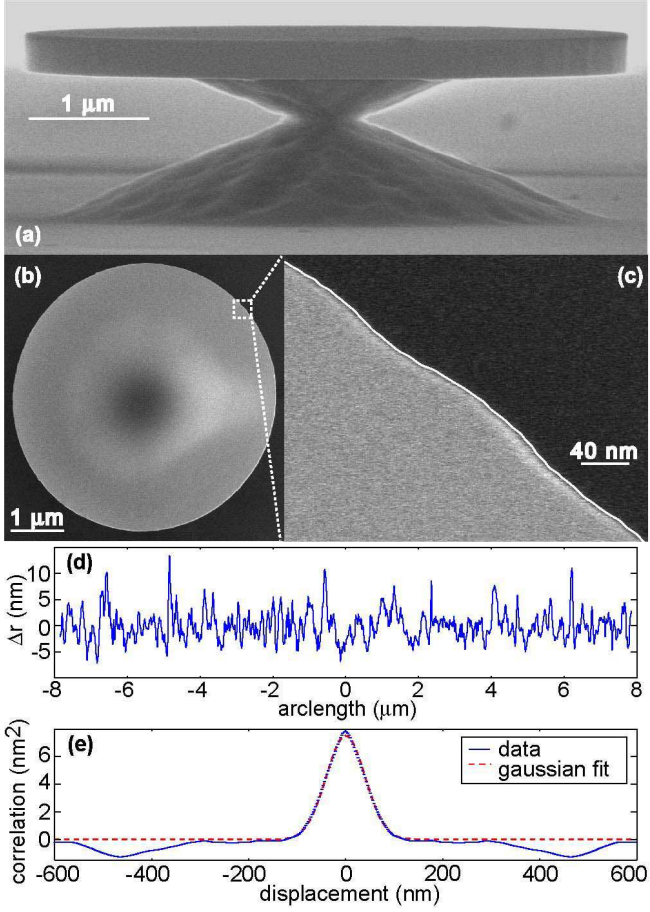, width=1.0\linewidth}
\caption{SEM micrographs of a $R=2.5$ $\mu$m Si microdisk: (a) side-view illustrating SiO$_{2}$ undercut and remaining pedestal, (b) high contrast top-view of disk, and (c) zoomed-in view of top edge showing disk-edge roughness and extracted contour (solid white line). (d) Plot of extracted contour versus arclength.  (e) Autocorrelation function of the microdisk contour and its Gaussian fit.} 
\label{SEM}
\end{center}
\end{figure}

The silicon microdisks in this work are fabricated from a silicon-on-insulator (SOI) wafer consisting of a $344$ nm thick p-doped Si layer of resistivity 1-3 $\Omega$-cm atop a two micron SiO$_{2}$ layer. Processing of the microdisks begins with the deposition of a 20 nm SiO$_{2}$ protective cap layer using plasma-enhanced chemical-vapor-deposition.  Electron beam lithography is used to create a polymer resist etch mask, and a low-bias voltage inductively-coupled-plasma reactive-ion-etch with SF$_{6}$:C$_{4}$F$_{8}$ gas chemistry\cite{Srinivasan7} then transfers the circular microdisk pattern into the top Si layer.  After dry-etching, the sample is immersed in buffered hydrofluoric acid to undercut the bottom SiO$_{2}$ cladding, as shown in Figure \ref{SEM}.  The thin $20$ nm SiO$_{2}$ top cap layer is also removed in this process, providing a clean, smooth top Si surface.  A final rinse in deionized water is performed, followed by a high-purity nitrogen spray drying step.

In order to characterize the microdisk resonators, an evanescent fiber taper coupling technique is employed\cite{Srinivasan7}.  In this process, an optical fiber is adiabatically drawn to a 1-2 $\mu$m diameter so that its evanescent field is made accessible to the environment.  In this work, the fiber taper is positioned to the side of the microdisks, with a center height equal to that of the middle of the microdisk.  Measurements of the taper transmission as a function of the lateral taper-microdisk gap ($g$) are then performed using a swept wavelength tunable laser source ($\lambda =$ 1509-1625 nm) with fine frequency resolution of $10$ MHz.  A set of paddle wheels are used to adjust the polarization state of the fiber taper mode in the microdisk coupling region, providing selective coupling to the TE-like (TM-like) WGMs with dominant electric field parallel (normal) to the plane of the microdisk.  For the 344 nm Si layer thickness of the microdisks studied here, only the fundamental vertical slab mode for TE and TM polarization are strongly guided.  As such, we only consider the fundamental vertical modes of the microdisks in what follows, labeling them simply by $p_{m,n}$, where $p$ is either TE or TM, and $m$ and $n$ are the characteristic radial and azimuthal number, respectively\cite{Little3}. 

Microdisks of two different sizes, radius $R=2.5$ and $4.5$ $\mu$m, are fabricated and tested.  A broad wavelength scan covering the 1509-1625 nm wavelength range is initially employed to map out the different microdisk modes.  As will be detailed elsewhere\cite{Borselli2}, the adjustable polarization state in the taper along with the WGMs' strength of coupling and linewidth is used to determine sets of modes with a common free spectral range.  These measurements provided a reliable determination of radial mode number.  An effective index two-dimensional model is then used to estimate the azimuthal number.  Using this mode identification technique we found that the highest $Q$ modes in both sizes of microdisks are consistently of TM polarization, and corresponded to the lowest radial number, $n \sim 1$.

\begin{figure}[t]
\begin{center}
\epsfig{figure=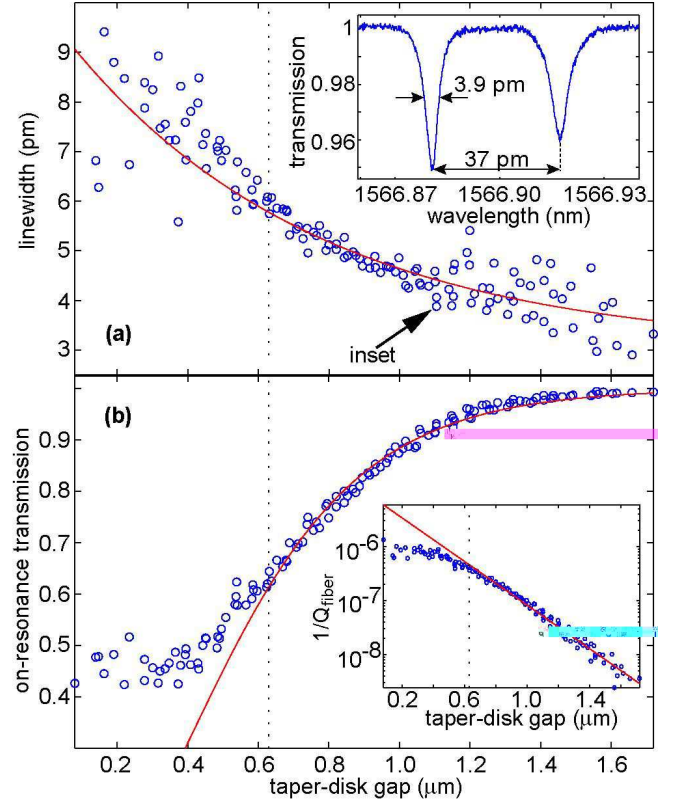, width=1.0\linewidth}
\caption{Fiber taper measurements of a $\text{TM}_{44,1}$ WGM of a microdisk with $R = 4.5$ $\mu$m.  (a)  Lorentzian full-width half-maximum (FWHM) linewidth versus taper-microdisk gap. (inset) Taper transmission showing high-$Q$ doublet.  (b) Resonant transmission depth versus taper-microdisk gap. (inset) loading versus taper-microdisk gap.}
\label{transmission}
\end{center}
\end{figure}

The inset of Fig. \ref{transmission}(a) shows the evanescent coupling to a $\text{TM}_{44,1}$ WGM of a $R=4.5$ $\mu$m microdisk, with tapered fiber positioned 1.1 $\mu$m laterally from the disk edge.  The observed double resonance dip (doublet) is a result of Rayleigh scattering from disk surface roughness as discussed below, which lifts the degeneracy of clockwise ($cw$) and counter-clockwise ($ccw$) propagating WGMs in the microdisk\cite{Weiss}.  Fitting the shorter wavelength mode of the doublet to a Lorentzian yields a loaded linewidth of $3.9$ pm with a $5\%$ coupling depth.  These measurements are repeated for varying taper-microdisk gaps and are recorded in Figure \ref{transmission}(a,b).  For $g>0.63$ $\mu$m, the data follows a two-port coupled mode theory with simple exponential loading dependence on taper-microdisk gap\cite{Cai}.  Fits based upon this model are shown as a solid line in each of the plots of Fig. \ref{transmission}.  The fiber loading of the microdisk is characterized here by a dimensionless effective quality factor, $Q_{\text{fiber}}$ (inset to Fig. \ref{transmission}(b)).  The asymptotic unloaded linewidth is found to be $3.0$ pm for this WGM, corresponding to a bare-cavity $Q$ of $5.2 \times 10^5$.  The observed non-exponential loading dependence for small $g$ is due to the large phase mismatch between the glass fiber and Si microdisk modes, further studies of which will be presented elsewhere\cite{Borselli2}.

\renewcommand{\arraystretch}{1.25}
\renewcommand{\extrarowheight}{0pt}

\begin{table}[t] \centering
\begin{tabularx}{\linewidth}{YYYY}
\hline\hline
\multirow{2}{\linewidth}{$\bigl(R(\mu\text{m}),p_{m,n}\bigr)$} & $V_{\text{eff,th.}}$  & $Q_{ss}$ & $\Delta\lambda_{\pm m}$ (pm) \\
                                                   & $(\lambda/n)^3$   & exp. (th.) & exp. (th.)     \\ 
\hline
$(2.5,\text{TM}_{23,1})$ & $5.3$ & $4.7$ $(3.5)\times 10^5$ &  $93$ $(112)$ \\ 
$(2.5,\text{TE}_{26,1})$ & $7.5$ & $9.0$ $(9.6)\times 10^4$ & $126$ $(146)$ \\
$(4.5,\text{TM}_{44,1})$ & $12.2$ & $5.2$ $(6.2)\times 10^5$ & $37$ $(46)$ \\ 
$(4.5,\text{TE}_{50,1})$ & $16.2$ & $1.7$ $(1.8)\times 10^5$ & $78$ $(59)$ \\
\hline
\hline
\end{tabularx}
\caption{Summary of theoretical and measured mode parameters for $R=2.5$ and $4.5$ $\mu$m Si microdisks.  Theoretical surface-scattering values for $Q_{ss}$ and $\Delta\lambda_{\pm m}$ are shown in parentheses.}
\label{table}
\end{table}%
\renewcommand{\arraystretch}{1.0}
\renewcommand{\extrarowheight}{0pt}

Similar measurements are performed for all $n \sim 1$ modes (of both polarizations) in each of the two different microdisk sizes, and a summary of the measured bare-cavity $Q$ and doublet mode-splitting values are given in Table \ref{table}.  In order to understand the limiting loss mechanisms and the observed trend of $Q$ with polarization and radial mode number, a complementary theoretical analysis is also performed.  Both the radiation limited $Q$ ($\gtrsim 10^{15}$\cite{Spillane_sid}) and the free carrier absorption $Q$ of the p-doped Si disk layer (estimated to be $>10^7$ from Ref. \cite{Soref1}), are far greater than the measured $Q$ values.  Surface state absorption is another possible loss mechanism; however, several trials are performed with differing final chemical cleaning steps with no perceivable change in $Q$.  The rather large measured splitting of the $cw$ and $ccw$ traveling wave modes indicates strong surface scattering\cite{Gorodetsky}, and therefore we believe the dominant source of optical loss to be scattering from index perturbations at the microdisk surface.

Surface roughness on the top and bottom surfaces of the microdisks analyzed in this work is negligible in comparison to the azimuthal variation of the disk radius [Fig. \ref{SEM}(c)], a result of mask erosion and deformation during the dry etching step.  For this type of roughness, the index perturbation can be approximated by $\delta\epsilon=\epsilon_{o}\delta n^{2}h\Delta r(s)\delta(r-R)\delta(z)$, where $\epsilon_{o}$ is the free space permittivity, $\delta n^{2}=n_{d}^{2}-n_{o}^{2}$, $n_{d} \sim 3.46$ is the Si refractive index, $n_{o} \sim 1$ is the index of the surrounding air, $h$ is the disk height, and $\Delta r(s)$ is the radial surface roughness relative to the unperturbed disk radius\cite{Little2}.  The parameters $r$, $s$, and $z$ correspond to the radius from the disk center, arc length along the disk perimeter, and height along the disk edge, respectively.  These perturbations of the disk radius set up polarization currents, $\vec{J}=-i\omega_{o}\delta\epsilon\vec{E}$, which couple the WGMs of the perfect microdisk to radiation modes or other nearly resonant WGMs.  The Volume Current Method\cite{Kuznetsov} can be used to calculate the radiated power into free space from $\vec{J}$, providing an estimate for the surface-scattering quality factor, $Q_{ss}$.  In the limit that the correlation length of the roughness ($L_{c}$) is much smaller than the wavelength in the material,

\begin{equation}
Q_{ss}=\frac{\lambda_{o}^{3}}{\pi^{5/2}n_{o}\bar{V}^{2}\sum_{\hat{e}}\bar{u}_{s}(\hat{e})G(\hat{e})},\label{Qss}%
\end{equation}

\noindent where $\bar{V}=\sqrt{RL_{c}}h\sigma_{r}\delta n^{2}$, $\bar{u}_{s}(\hat{e})$ is the average $\hat{e}$-polarized electric field energy density at the disk edge normalized to unit modal energy, $\sigma_{r}$ is the standard deviation of the roughness, and $G(\hat {e})=\{4/3,2,2/3\}$ is a geometrical radiation factor for the $\hat{e}=\{\hat{r},\hat{\phi},\hat{z}\}$ different electric field polarizations.  The mode coupling between counterpropagating $cw$ and $ccw$ traveling wave modes can also be found via a time-dependent perturbation theory\cite{Gorodetsky}.  This coupling breaks the degeneracy of the $cw$ and $ccw$ modes ($\pm m$), creating two standing wave modes with wavelength splitting

\begin{equation}
\frac{\Delta\lambda_{\pm m}}{\lambda _{o}}=\left( \frac{\pi}{4}\right)^{3/4}\bar{V}\bar{u}_{s}\exp\left[ -\frac{1}{2}\left(\frac{L_{c}m}{R}\right)^{2}\right].\label{doublet}%
\end{equation}
\noindent where $\bar{u}_{s}= \sum_{\hat{e}}\bar{u}_{s}(\hat{e})$.  SEM micrographs such as those shown in Figs. \ref{SEM}(b,c) provide high contrast images from which edge contours with near nanometer resolution can be obtained [Fig. \ref{SEM}(d)].  From a Gaussian fit to the autocorrelation function of the surface roughness [Fig. \ref{SEM}(e)], estimates for $\sigma_{r}$ and $L_{c}$ are found to be $2.8$ nm and $40$ nm, respectively, typical of the fabricated microdisks in this work.  Using the results from this roughness analysis in Eqs. (\ref{Qss}) and (\ref{doublet}), theoretical estimates for $Q$ and $\Delta\lambda_{\pm m}$ are calculated and tabulated in Table \ref{table}, with good agreement found between theory and the measured microdisk properties.  In these calculations, an effective index model based on Ref. \cite{Little3} is used to estimate $\bar{u}_{s}(\hat{e})$ and the effective mode volume $V_{\text{eff}}$\cite{note_mode_volume}, with the approximate TM modes given by a single $\hat{z}$ field component and the TE modes consisting of $\hat{r}$ and $\hat{\phi}$ electric field components.  The salient result of this analysis is that a TE mode has roughly three times the radiated power versus a comparable TM mode.  This is explained by the fact that both the $\hat{r}$ and $\hat{\phi}$ components couple to radiation modes more strongly than the $\hat{z}$ component of the field.  The modes of higher $n$ number are also measured and found to be lower $Q$ as predicted by the disk-edge surface-scattering theory, although for $n \geqslant 2$ the effects of the underlying SiO$_{2}$ pedestal may be significant.

In summary, Si microdisks of a few microns in radius have been fabricated with measured $Q > 5 \times 10^5$.  In contrast to theoretical predictions for an ideal microdisk, we find that the TM modes have significantly higher measured $Q$ values than TE modes due to their inherent reduced sensitivity to disk-edge surface roughness.  In comparison to previously measured high-$Q$ semiconductor microdisks \cite{Gayral}, the $Q$ of the smallest WGMs measured in this work are $> 20$ times larger while maintaining a similar effective mode volume.  

The authors thank S. Spillane, T. Johnson, and H. Huang for helpful contributions to this work.  M.B. thanks the Moore Foundation, NPSC, and HRL Laboratories, and K.S. thanks the Hertz Foundation for their graduate fellowship support.

\bibliographystyle{apsrev}
\bibliography{./SID}

\begin{thebibliography}{20}
\bibitem[{not()}]{note_mode_volume}
\bibinfo{note}{$V_{\text{eff}}$ is defined as the electric field energy within
  the cavity normalized to the peak electric field energy
  density\cite{Andreani}.}
\end{thebibliography}
\end{document}